\documentclass[twocolumn,eqsecnum,amsmath,amssymb,nofootinbib]{revtex4}
\usepackage{graphicx}
\usepackage{dcolumn}
\usepackage{bm}

\begin{document}

\title{
\hfill{\normalsize\vbox{\hbox{\rm DPNU-04-14}  }}\\
\vspace{0.1cm}
Scalar Mesons in Radiative Decays and
$\pi$-$\pi$ Scattering~\footnote{%
Talk given at YITP workshop on
``Multi-quark Hadrons; four, five and more ?",
February 17-19, 2004, Yukawa Insitute, Kyoto, Japan.}
}

\author{Masayasu Harada}

\affiliation{Department of Physics, Nagoya University, 
  Nagoya, 464-8602, Japan}

\begin{abstract}
In this write-up,
I summarize the analyses on the low-lying scalar mesons
I have done recently with my collaborators.
I first briefly review the previous analyses on the hadronic
processes related to the scalar mesons, which shows that
the scalar nonet takes dominantly
the $qq\bar{q}\bar{q}$ structure.
Next, I summarize our analysis on the radiative decays involving
the scalar mesons, which indicates that it is difficult to
distinguish $qq\bar{q}\bar{q}$ picture and $q\bar{q}$ picture
just from radiative decays.
Finally, I summarize our recent analysis on the $\pi$-$\pi$ scattering
in the large $N_c$ QCD, which indicates that the $\sigma$ meson is
likely the $qq\bar{q}\bar{q}$ state.
\end{abstract}

\maketitle


\section{I\lowercase{ntroduction}}

According to recent theoretical and experimental analyses, there
is a possibility that nine light scalar mesons exist below 1 GeV, and
they form a scalar nonet~\cite{proceedings}.
In addition to the well established $f_0(980)$ and
$a_0(980)$ evidence of both experimental and theoretical 
nature for a very broad 
$\sigma$ ($\simeq560$)
and  a very broad $\kappa$ 
($\simeq900$)
has been presented.

As is stressed 
in Ref.~\cite{Black-Fariborz-Sannino-Schechter:99},
the masses of the above low-lying scalar mesons
do not obey the ``ideal mixing'' pattern
which nicely explains the masses of mesons made from a quark
and an anti-quark such as vector mesons~\cite{Okubo}.
As is shown 
in Ref.~\cite{Black-Fariborz-Sannino-Schechter:99},
the ``ideal mixing'' pattern qualitatively
explains the mass hierarchy of the
scalar nonet when the members of the nonet have
a $qq\bar{q}\bar{q}$ quark structure 
proposed in
Ref.~\cite{Jaffe}.
In this 4-quark picture,
two quarks are combined to make a diquark which
together with an anti-diquark
forms a scalar meson.
The resultant scalar mesons have the same quantum numbers as
the ordinary scalar mesons made from the quark and 
anti-quark (2-quark picture).
It is difficult to clarify the quark structure
of the low-lying scalar mesons just from their quantum numbers.
The patterns of the interactions of the scalar mesons to other mesons
made from $q\bar{q}$, on the other hand, depend on the quark structure
of the scalar mesons.
I expect that the analysis on the interactions of the scalar mesons
will shed some lights on the quark structure of the scalar nonet.
Actually, in 
Refs.~\cite{Black-Fariborz-Sannino-Schechter:99,Fariborz-Schechter},
several hadronic processes related to the scalar mesons are studied.
They concluded that the scalar nonet takes dominantly the
$qq\bar{q}\bar{q}$ structure.

Recently, for getting more informations on the structure
of the low-lying scalar mesons,
we studied the radiative decays involving scalar mesons~\cite{BHS}
and the $\pi$-$\pi$ scattering in the large $N_c$ 
QCD~\cite{Harada-Sannino-Schechter:04}.
In this write-up
I will summarize these analyses, especially focusing on 
the quark structure of the low-lying scalar nonet,
and show how these processes give a clue for understanding 
the structure of the scalar mesons.

This write-up is organized as follows:
In section~\ref{sec:ELSM}, following 
Refs.~\cite{Black-Fariborz-Sannino-Schechter:99,Fariborz-Schechter},
I will briefly review the analyses
on the hadronic processes related to the scalar mesons. 
Next, in section~\ref{sec:rad}, I will briefly summarize
the analysis on the radiative decays involving the scalar 
mesons based on the 4-quark picture~\cite{BHS}.
I also present a new result on the analysis on the decay processes
based on the 2-quark picture~\cite{BHS:prep}.
In section~\ref{sec:pps}, I will summarize our analysis on 
the $\pi$-$\pi$ scattering in the large $N_c$ 
QCD~\cite{Harada-Sannino-Schechter:04}.
Finally, in section~\ref{sec:sum}, I will give a brief summary.

\section{E\lowercase{ffective} L\lowercase{agrangian for}
S\lowercase{calar} M\lowercase{esons} }
\label{sec:ELSM}

In this section
I briefly review previous 
analyses~\cite{Black-Fariborz-Sannino-Schechter:99,Fariborz-Schechter}
on the masses
of scalar mesons and hadronic processes related to the
scalar mesons. 

In Ref.~\cite{Black-Fariborz-Sannino-Schechter:99},
the scalar meson nonet is embedded into the $3\times3$ matrix
field $N$ as
\begin{equation}
N = 
\left(\begin{array}{ccc}
   \displaystyle \left(N_T+a_0^0\right) / \sqrt{2}
 & \displaystyle a_0^+
 & \displaystyle \kappa^+
\\
   \displaystyle a_0^-
 & \displaystyle \left(N_T-a_0^0\right) / \sqrt{2}
 & \displaystyle \kappa^0
\\
   \displaystyle \kappa^-
 & \displaystyle \bar{\kappa}^0
 & \displaystyle N_S
\\
\end{array}\right)
\ ,
\label{scalar nonet}
\end{equation}
where
$N_T$ and $N_S$ represents the ``ideally mixed'' fields.
The physical $\sigma(560)$ and $f_0(980)$ fields are expressed by
the linear combinations of these $N_T$ and $N_S$ as
\begin{equation}
\left(\begin{array}{c}
  \displaystyle \sigma \\ \displaystyle f_0 \\
\end{array}\right)
=
\left(\begin{array}{cc}
 \displaystyle \cos\theta_S & \displaystyle -\sin\theta_S \\ 
 \displaystyle \sin\theta_S & \displaystyle \cos\theta_S \\
\end{array}\right)
\,
\left(\begin{array}{c}
  \displaystyle N_S \\ \displaystyle N_T \\
\end{array}\right)
\ ,
\label{mixing}
\end{equation}
where $\theta_S$ is the scalar mixing angle.

The scalar mixing angle $\theta_S$
can parameterize the quark contents
of the scalar nonet field:
When $\theta_S = \pm 90^\circ$,
the $\sigma$ and $f_0$ fields are embedded into the nonet field
as
\begin{equation}
N = 
\left(\begin{array}{ccc}
   \displaystyle \left(\sigma+a_0^0\right) / \sqrt{2}
 & \displaystyle a_0^+
 & \displaystyle \kappa^+
\\
   \displaystyle a_0^-
 & \displaystyle \left(\sigma-a_0^0\right) / \sqrt{2}
 & \displaystyle \kappa^0
\\
   \displaystyle \kappa^-
 & \displaystyle \bar{\kappa}^0
 & \displaystyle f_0
\\
\end{array}\right)
\ .
\end{equation}
This is a natural assignment of scalar meson nonet based on 
the $q\bar{q}$ picture:
\begin{equation}
\sim \ 
\left(\begin{array}{ccc}
   \displaystyle \bar{u} u 
 & \displaystyle \bar{d} u
 & \displaystyle \bar{s} u
\\
   \displaystyle \bar{u} d
 & \displaystyle \bar{d} d
 & \displaystyle \bar{s} d
\\
   \displaystyle \bar{u} s
 & \displaystyle \bar{u} s
 & \displaystyle \bar{s} s
\\
\end{array}\right)
\ .
\end{equation}
On the other hand, when $\theta_S = 0^\circ$ or $180^\circ$,
the scalar nonet field $N$ becomes
\begin{equation}
N = 
\left(\begin{array}{ccc}
   \displaystyle \left(f_0+a_0^0\right) / \sqrt{2}
 & \displaystyle a_0^+
 & \displaystyle \kappa^+
\\
   \displaystyle a_0^-
 & \displaystyle \left(f_0-a_0^0\right) / \sqrt{2}
 & \displaystyle \kappa^0
\\
   \displaystyle \kappa^-
 & \displaystyle \bar{\kappa}^0
 & \displaystyle \sigma
\\
\end{array}\right)
\ ,
\end{equation}
which is a natural assignment of scalar meson nonet based on 
the $qq\bar{q}\bar{q}$ picture:
\begin{equation}
\sim \ 
\left(\begin{array}{ccc}
   \displaystyle \bar{s} \bar{d} d s
 & \displaystyle \bar{s} \bar{d} u s
 & \displaystyle \bar{s} \bar{d} u d
\\
   \displaystyle \bar{s} \bar{u} d s
 & \displaystyle \bar{s} \bar{u} u s
 & \displaystyle \bar{s} \bar{u} u d
\\
   \displaystyle \bar{u} \bar{d} d s
 & \displaystyle \bar{u} \bar{d} u s
 & \displaystyle \bar{u} \bar{d} u d
\\
\end{array}\right)
\ .
\end{equation}
Then, the present treatment of nonet field with the scalar mixing
angle can express both pictures for quark contents.

By using the scalar nonet field introduced above,
the effective Lagrangian for the scalar meson masses
are expressed as~\cite{Black-Fariborz-Sannino-Schechter:99}
\begin{eqnarray}
{\mathcal L}_{\rm mass} &=&
  - a \, \mbox{tr} \left[ N N \right]
  - b \, \mbox{tr} \left[ {\mathcal M} N N \right]
  - c \, \mbox{tr} \left[ N \right] \, \mbox{tr} \left[ N \right]
\nonumber\\
&& {}
  - d \, \mbox{tr} \left[ {\mathcal M} N \right]
      \, \mbox{tr} \left[ N \right]
\ ,
\end{eqnarray}
where $a$, $b$, $c$ and $d$ are real constants, and ${\mathcal M}$
is the ``spurion matrix'' expressing the explicit chiral 
symmetry breaking due to the current quark masses.
This ${\mathcal M}$ is defined by 
${\mathcal M} = \mbox{diag}(1,1,x)$,
where $x$ is the ratio of strange to non-strange quark masses with the
isospin invariance assumed.
Note that the scalar mixing angle is expressed by a combination of
the parameters
$a$, $b$, $c$ and $d$.

Here I 
use the following values of the masses of the
scalar nonet as inputs:
\begin{equation}
M_{a_0} \simeq 980\,\mbox{MeV} \ ,
\quad
M_{f_0} \simeq 980\,\mbox{MeV} \ ,
\end{equation}
listed in Particle Data Group (PDG) table~\cite{PDG00,PDG},
\begin{equation}
M_\sigma \simeq 560 \,\mbox{MeV}\ ,
\end{equation}
determined from the $\pi$-$\pi$ 
scattering~\cite{Harada-Sannino-Schechter:96},
and 
\begin{equation}
M_{\kappa} \simeq 900 \,\mbox{MeV} \ ,
\end{equation}
determined from the $\pi$-$K$ scattering~\cite{BFSS-piK}.
The above choice yields the two possible solutions for 
the scalar mixing angle~\cite{Black-Fariborz-Sannino-Schechter:99}
\begin{eqnarray}
\theta_S \sim -20^\circ \ , 
\label{sol 20}
\\
\theta_S \sim -90^\circ \ .
\label{sol 90}
\end{eqnarray}
Solution in Eq.~(\ref{sol 20}) corresponds to the case where
the scalar nonet is dominantly made from $qq\bar{q}\bar{q}$, while
solution in Eq.~(\ref{sol 90}) to the case
where it is from $q\bar{q}$.

For determining the scalar mixing angle,
the authors of Ref.~\cite{Black-Fariborz-Sannino-Schechter:99}
considered the tri-linear scalar-pseudoscalar-pseudoscalar
interaction.
There the pseudoscalar mesons are embedded
into the nonet field as
\begin{equation}
P = \left(\begin{array}{ccc}
   \displaystyle \left(\eta_T+\pi^0\right)/\sqrt{2}
 & \displaystyle \pi^+
 & \displaystyle K^+
\\
   \displaystyle \pi^-
 & \displaystyle \left(\eta_T-\pi^0\right)/\sqrt{2}
 &  \displaystyle K^0
\\
   \displaystyle K^-
 & \displaystyle \bar{K}^0
 & \displaystyle \eta_S
\end{array}\right)
\ ,
\end{equation}
where $\eta_T$ and $\eta_S$ denote the ideally mixed fields.
Based on the two-mixing-angle scheme introduced in 
Ref.~\cite{SSW}
the physical $\eta$ and $\eta'$ fields are expressed by
the linear combinations of $\eta_T$ and $\eta_S$.

By using the scalar meson nonet field $N$ defined in 
Eq.~(\ref{scalar nonet}) together with the above pseudoscalar
nonet field $P$, the general SU(3) flavor invariant 
scalar-pseudoscalar-pseudoscalar interaction is written 
as~\cite{Black-Fariborz-Sannino-Schechter:99}
\begin{eqnarray}
\lefteqn{
 - {\mathcal L}_{NPP} 
 =
 A \epsilon^{abc} \epsilon_{def} \, N_a^d \,
  \partial_\mu P_b^e \, \partial^\mu P_c^f 
}
\nonumber\\
&& {}
+ B \, \mbox{tr}[ N ] \, \mbox{tr} [ \partial_\mu P \partial^\mu P ]
+ C \mbox{tr} [ N \partial_\mu P ] \, \mbox{tr} [ \partial^\mu P ]
\nonumber\\
&& {}
+ D \mbox{tr} [N] \,\mbox{tr}[\partial_\mu P]
  \, \mbox{tr} [ \partial^\mu P ]
\ ,
\label{NPP Lag}
\end{eqnarray}
where $A$, $B$, $C$ and $D$ are four real constants,
and $a,b,c=1,2,3$ denote flavor indices.
The derivatives
of the pseudoscalars were introduced in order that 
Eq.~(\ref{NPP Lag}) properly follows from a chiral invariant
Lagrangian in which the field $P$ transforms non-linearly under
chiral transformation.

In Refs.~\cite{Black-Fariborz-Sannino-Schechter:99,%
Fariborz-Schechter}, four parameters $A$, $B$, $C$, $D$ and
the scalar mixing angle are determined by fitting them
to the experimental data of the $\pi$-$K$ scattering and 
the $\eta^\prime \rightarrow \eta\pi\pi$ decay together
with the $\pi$-$\pi$ scattering.
The resultant best fitted values for $A$, $B$, $C$ and $D$ are
\begin{eqnarray}
&&
 A \simeq 2.5 \,\mbox{GeV}^{-1} \ , \quad
 B \simeq -2.0 \,\mbox{GeV}^{-1} \ , 
\nonumber\\
&&
 C \simeq -2.3 \,\mbox{GeV}^{-1} \ , \quad
 D \simeq -2.3 \,\mbox{GeV}^{-1} \ .
\end{eqnarray}
The best fitted value of the scalar mixing angle is
\begin{equation}
\theta_S \simeq -20^\circ \ ,
\end{equation}
which implies that the scalar meson takes dominantly the
$qq\bar{q}\bar{q}$ structure.
It should be noticed that the coupling constant of the
$f_0$-$\pi$-$\pi$ interaction determined from the $\pi$-$\pi$
scattering~\cite{Harada-Sannino-Schechter:96} plays an
important role to constrain the value of the mixing angle.

\section{R\lowercase{adiative} D\lowercase{ecays}
 I\lowercase{nvolving} S\lowercase{calar} M\lowercase{esons}
\label{sec:rad}
}

In the previous section, I briefly reviewed the analyses
done in Refs.~\cite{Black-Fariborz-Sannino-Schechter:99,%
Fariborz-Schechter}, which shows that the experimental data
of the hadronic decay processes involving scalar mesons
give $\theta_S \simeq -20^\circ$,
i.e., the scalar meson is dominantly made from $qq\bar{q}\bar{q}$.
In this section, I show our analysis on the radiative
decays involving the scalar mesons done in Ref.~\cite{BHS}.

\begin{figure}[htbp]
\includegraphics [width = 6.5cm] {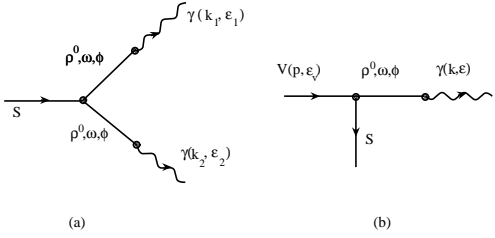}
\caption{%
Feynman diagrams for (a)~$S\rightarrow \gamma\gamma$ and
(b)~$V \rightarrow S \gamma$.
}\label{fig:1}
\end{figure}

In Ref.~\cite{BHS}, 
the trilinear scalar-vector-vector terms were included
into the effective Lagrangian as
\begin{eqnarray}
&&
{\cal L}_{SVV} =  \beta_A \,
\epsilon_{abc} \epsilon^{a'b'c'}
\left[ F_{\mu\nu}(\rho) \right]_{a'}^a
\left[ F_{\mu\nu}(\rho) \right]_{b'}^b N_{c'}^c
\nonumber\\
&& \quad
{}+
 \beta_B \, \mbox{tr} \left[ N \right]
\mbox{tr} \left[ F_{\mu\nu}(\rho) F_{\mu\nu}(\rho) \right]
\nonumber\\
&& \quad
{}+
 \beta_C \, \mbox{tr} \left[ N F_{\mu\nu}(\rho) \right]
\mbox{tr} \left[ F_{\mu\nu}(\rho) \right]
\nonumber\\
&& \quad
{}+
 \beta_D \, \mbox{tr} \left[ N \right]
\mbox{tr} \left[ F_{\mu\nu}(\rho) \right]
\mbox{tr} \left[ F_{\mu\nu}(\rho) \right]
\ .
\label{SVV}
\end{eqnarray}
where $N$ is the scalar nonet field defined in 
Eq.~(\ref{scalar nonet}).
$F_{\mu\nu}(\rho)$ is the field strength of the
vector meson fields defined as
\begin{eqnarray}
&&
  F_{\mu\nu}(\rho) = 
  \partial_\mu \rho_\nu - \partial_\nu \rho_\mu - i 
  \tilde{g} \left[ \rho_\mu \,,\, \rho_\nu \right]
\ ,
\label{transf}
\end{eqnarray}
where $\tilde{g} \simeq4.04$~\cite{Harada-Schechter} is
the coupling constant.
(A term 
$\sim \mbox{tr}( FFN )$ is linearly dependent on the four shown).
In Ref.~\cite{BHS}, the vector meson dominance is assumed to be
satisfied in the radiative decays involving the scalar mesons.  Then,
the above Lagrangian (\ref{SVV}) determines all
the relevant interactions.
Actually, the $\beta_D$ term will not contribute so there
are only three relevant parameters $\beta_A$, $\beta_B$ and $\beta_C$.
Equation~(\ref{SVV}) is analogous to the $PVV$ 
interaction~\footnote{%
 It was shown~\cite{BKY:88-HY:03} that the complete vector
 meson dominance (VMD) is violated in the 
 $\omega\rightarrow \pi^0 \pi^+ \pi^-$ decay which
 is expressed
 by $PVV$ interactions.  However, since the VMD is satisfied
 in other processes such $\pi^0 \rightarrow \gamma \gamma^\ast$
 as well as
 in the electromagnetic form factor of pion, 
 it is assumed to be held
 in the processes related to $SVV$ interactions
 in Ref.~\cite{BHS}.
}
which was
originally introduced as a $\pi\rho\omega$ coupling a long time 
ago~\cite{GSW}. 
One can now compute the amplitudes for 
$S\rightarrow\gamma\gamma$ and $V \rightarrow S \gamma$ according to
the diagrams of Fig.~\ref{fig:1}.

The decay matrix element for $S \rightarrow \gamma \gamma$ is
written as 
$(e^2/\tilde{g}^2) X_S \times \bigl(
k_1\cdot k_2 \, \epsilon_1\cdot \epsilon_2 -
k_1\cdot \epsilon_2 \, k_2 \cdot \epsilon_1
\bigr)$ 
where $\epsilon_\mu$ stands for the photon polarization vector.  It is
related to the width by
\begin{equation}
\Gamma \left( S \rightarrow \gamma\gamma \right)
=
\alpha^2 \frac{\pi}{4} m_S^3
\left\vert \frac{X_S}{\tilde{g}^2} \right\vert^2
\ ,
\label{swidth}
\end{equation}
where $X_S$ takes on the specific forms:
\begin{eqnarray}
X_{\sigma} &=&
  \frac{4}{9} \beta_A 
    \left( \sqrt{2} s - 4 c \right)
  {} + \frac{8}{3} \beta_B
    \left( c - \sqrt{2} s \right)
\ ,
\nonumber\\
X_{f_0} &=&
  - \frac{4}{9} \beta_A 
    \left( \sqrt{2} c + 4 s \right)
  {} + \frac{8}{3} \beta_B
    \left( \sqrt{2} c + s \right)
\ , 
\nonumber\\
X_{a_0} &=& \frac{4\sqrt{2}}{3} \beta_A 
\ .
\label{samps}
\end{eqnarray}
In the above expressions
$\alpha=e^2/(4\pi)$, $s = \sin \theta_S$ and $c = \cos\theta_S$
where 
the scalar
mixing angle, $\theta_S$ is defined in Eq.~(\ref{mixing}).
Furthermore, ideal mixing for the vector mesons, with
$\rho^0 = (\rho_1^1 - \rho_2^2)/\sqrt{2}$,
$\omega = (\rho_1^1 + \rho_2^2)/\sqrt{2}$,
$\phi = \rho_3^3$, was assumed for simplicity.

Similarly to the one for $S \rightarrow \gamma\gamma$, 
the decay matrix element for $V\rightarrow S \gamma$ is
written as $(e/\tilde{g}) C_V^S \times \left[
p\cdot k \epsilon_V \cdot \epsilon 
- p \cdot \epsilon k \cdot \epsilon_V
\right]$.
It is related to the width by
\begin{equation}
\Gamma (V \rightarrow S \gamma ) =
\frac{\alpha}{3} \left\vert k_V^S \right\vert^3
\left\vert \frac{C_V^S}{\tilde{g}} \right\vert^2
\ ,
\end{equation}
where $k_V^S = (m_V^2 - m_S^2)/(2m_V)$ is the photon momentum in the
$V$ rest frame.
For the energetically allowed $V \rightarrow S \gamma$ processes we
have
\begin{eqnarray}
C_\phi^{f_0} &=& 
 \frac{2\sqrt{2}}{3} \beta_A c
 - \frac{4}{3} \beta_B \left( \sqrt{2} c + s \right)
\nonumber\\
&& \quad
 + \frac{\sqrt{2}}{3} \beta_C
   \left( c - \sqrt{2} s \right)
\ , 
\nonumber\\
C_\phi^{\sigma} &=& 
 - \frac{2\sqrt{2}}{3} \beta_A s
 - \frac{4}{3} \beta_B 
   \left( c - \sqrt{2} s \right)
\nonumber\\
&& \quad
 - \frac{2}{3} \beta_C
   \left( c + \frac{1}{\sqrt{2}} s \right)
\ , 
\nonumber\\
C_\phi^{a_0} &=& \sqrt{2} \left( \beta_C - 2 \beta_A \right)\ , 
\nonumber\\
C_\omega^{\sigma} &=& 
 \frac{2\sqrt{2}}{3} \beta_A 
   \left( c + \sqrt{2} s \right)
 + \frac{2\sqrt{2}}{3} \beta_B 
   \left( c - \sqrt{2} s \right)
\nonumber\\
&& \quad
 - \frac{2}{3} \beta_C
   \left( \sqrt{2} c + s \right),
\nonumber\\
C_{\rho^0}^\sigma &=& -2\sqrt{2} \beta_A c + 2 \sqrt{2} \beta_B
\left( c- \sqrt{2} s \right).
\label{vamps}
\end{eqnarray}

In addition, the same model predicts amplitudes for
 the energetically allowed
$S\rightarrow V \gamma$ processes:
$f_0 \rightarrow \omega \gamma$, $f_0 \rightarrow \rho^0 \gamma$,
$a_0^0 \rightarrow \omega \gamma$,
$a_0^0 \rightarrow \rho^0 \gamma$ and,
if $\kappa^0$ is sufficiently heavy 
$\kappa^0 \rightarrow K^{\ast0} \gamma$.
The corresponding width is
\begin{equation}
\Gamma (S \rightarrow V \gamma ) =
 \alpha \left\vert k_S^V \right\vert^3
\left\vert \frac{D_S^V}{\tilde{g}} \right\vert^2
\ ,
\end{equation}
where $k_S^V = (m_S^2 - m_V^2)/(2m_S)$ and
\begin{eqnarray}
D_{f_0}^\omega &=& 
 \frac{2}{3} \beta_A \left( - 2 c +\sqrt{2} s \right)
 + \frac{2}{3} \beta_B \left( 2 c +\sqrt{2} s \right)
\nonumber\\
&& \quad
 + \frac{2}{3} \beta_C
   \left( c - \sqrt{2} s \right)
\ , 
\nonumber\\
D_{f_0}^{\rho^0} &=& 
 - 2 \sqrt{2} \beta_A s
 + 2 \beta_B 
   \left(2 c +\sqrt{2} s \right)
\ , 
\nonumber\\
D_{a_0}^\omega &=& 2 \beta_C \ ,
\nonumber\\
D_{a_0}^{\rho^0} &=&  \frac{4}{3} \beta_A \ ,
\nonumber\\
D_{\kappa^0}^{K^{\ast0}} &=& 
 - \frac{8}{3} \beta_A 
\ .
\label{svgamps}
\end{eqnarray}

Let me show the results obtained in 
Ref.~\cite{BHS} together with new results from a recent
analysis~\cite{BHS:prep}.
I should stress again that
all the different decay amplitudes are described by 
only three parameters
$\beta_A$, $\beta_B$ and $\beta_C$.

In Ref.~\cite{BHS}, the value of $\beta_A$ was determined
from the $a_0 \rightarrow \gamma\gamma$ process.
Substituting 
$\Gamma_{\rm exp}(a_0\rightarrow \gamma\gamma) =
(0.28\pm0.09)\,\mbox{keV}$ (obtained using \cite{PDG}
$B(a_0 \rightarrow K \bar{K})/B( a_0 \rightarrow \eta \pi) =
0.177 \pm 0.024$)
into Eqs.~(\ref{swidth}) and (\ref{samps})
yields
\begin{equation}
\beta_A = (0.72\pm0.12)\,\mbox{GeV} \ ,
\label{valA}
\end{equation}
where positive in sign was assumed.
By using this value, 
the value of $\beta_C$ is determined 
from 
$\Gamma_{\rm exp}(\phi \rightarrow a_0 \gamma) =
(0.47\pm0.07)\,\mbox{keV}$ 
(obtained by assuming $\phi\rightarrow
\eta\pi^0\gamma$ is dominated by $\phi\rightarrow a_0\gamma$)
and Eq.~(\ref{vamps}) 
as
\begin{equation}
\beta_C = 
(7.7\pm0.5 \,,\, -4.8\pm0.5)\,\mbox{GeV}^{-1} \ .
\label{valC}
\end{equation}
It should be stressed that 
the values of $\beta_A$ and $\beta_C$ obtained above 
are independent of the mixing angle $\theta_S$,
and that 
$\vert\beta_A\vert$ is almost an order of
magnitude smaller than $\vert\beta_C\vert$.
As one can see from Eq.~(\ref{svgamps}),
the amplitude $D_{a_0}^\omega$ is given by $\beta_C$ while
$D_{a_0}^{\rho^0}$ is given by only $\beta_A$.
Then, the large hierarchy between $\beta_C$ and $\beta_A$ implies
that there is a large hierarchy between 
$\Gamma(a_0\rightarrow\omega\gamma)$ and 
$\Gamma(a_0\rightarrow\rho\gamma)$.
Actually, by using the values of $\beta_A$ and $\beta_C$ given
in Eqs.~(\ref{valA}) and (\ref{valC}),
they are estimated as
\begin{eqnarray}
&& \Gamma(a_0\rightarrow\omega\gamma) 
  = \left( 641\pm87\,,\,251\pm54 \right) \,\mbox{keV} \ ,
\nonumber\\
&& \Gamma(a_0\rightarrow\rho\gamma) 
  =  3.0\pm1.0 \,\mbox{keV} \ .
\end{eqnarray}
This implies that there is a large hierarchy between
$\Gamma(a_0\rightarrow\omega\gamma)$ and
$\Gamma(a_0\rightarrow\rho\gamma)$
which is caused by an order of magnitude difference between
$\vert\beta_C\vert$ and $\vert\beta_A\vert$.

I next show how to determine the value of $\beta_B$ from 
the $f_0\rightarrow \gamma\gamma$ process.
$X_{f_0}$ in Eq.~(\ref{samps}) depends on
$\beta_B$ as well as on $\beta_A$ and the 
scalar mixing angle $\theta_S$.
Here the scalar mixing angle $\theta_S$ is taken as 
\begin{equation}
\theta_S \simeq - 20^{\circ}\ ,
\end{equation}
which is characteristic of $qq{\bar q}{\bar q}$ type 
scalars~\cite{Black-Fariborz-Sannino-Schechter:99}.
By using this and the value of $\beta_A$ in Eq.~(\ref{valA}),
$\Gamma_{\rm exp}(f_0 \rightarrow \gamma\gamma) = 
0.39\pm0.13\,\mbox{keV}$ yields
\begin{equation}
\beta_B = (0.61\pm0.10 \,,\,-0.62\pm0.10)\,\mbox{GeV}^{-1} \ .
\end{equation}
This implies that $\vert\beta_B\vert$ is on the order of 
$\vert\beta_A\vert$, and almost an order of magnitude smaller
than $\vert\beta_C\vert$.
Equation~(\ref{svgamps}) shows that
$D_{f_0}^\omega$ includes $\beta_C$ while
$D_{f_0}^\rho$ does not.
Thus, there is a large hierarchy between 
decay widths of
$f_0\rightarrow\omega\gamma$ and
$f_0\rightarrow\rho\gamma$:
The typical predictions are given by
\begin{eqnarray}
&& \Gamma(f_0\rightarrow\omega\gamma) = ( 88\pm17)\,\mbox{keV}
\ ,
\nonumber\\
&& \Gamma(f_0\rightarrow\rho\gamma) = ( 3.3\pm2.0)\,\mbox{keV}
\ .
\label{pred f0}
\end{eqnarray}
This implies that there is a large hierarchy between
$\Gamma(f_0\rightarrow\omega\gamma)$ and
$\Gamma(f_0\rightarrow\rho\gamma)$ which is 
caused by the fact that
$\vert\beta_C\vert$ is 
an order of magnitude larger than
$\vert\beta_A\vert$ and $\vert\beta_B\vert$.
I summarize the fitted values
$\beta_A$, $\beta_B$ and $\beta_C$
together with several predicted values of 
the decay widths of 
$V \rightarrow S + \gamma$ and $S \rightarrow V + \gamma$
in Table~\ref{tab1}.
\begin{table}[htbp]
\begin{center}
\begin{tabular}{lrr}
\hline
$\beta_A$ & $0.72\pm0.12$ & $0.72\pm0.12$ \\
$\beta_B$ & $0.61\pm0.10$ & $-0.62\pm0.10$ \\
$\beta_C$ & $7.7\pm0.52$ & $7.7\pm0.52$ \\
\hline
$\Gamma(\sigma\rightarrow\gamma\gamma)$ 
   & $0.024\pm0.023$ & $0.38\pm0.09$ \\
$\Gamma(\phi\rightarrow\sigma\gamma)$ & $137\pm19$ & $33\pm9$ \\
$\Gamma(\omega\rightarrow\sigma\gamma)$ & $16\pm3$ & $33\pm4$ \\
$\Gamma(\rho\rightarrow\sigma\gamma)$ & $0.23\pm0.47$ & $17\pm4$ \\
$\Gamma(f_0\rightarrow\omega\gamma)$ 
  & $126\pm20$ & $88\pm17$ \\
$\Gamma(f_0\rightarrow\rho\gamma)$ & $19\pm5$ & $3.3\pm2.0$ \\
$\Gamma(a_0\rightarrow\omega\gamma)$ & $641\pm87$ & $641\pm87$ \\
$\Gamma(a_0\rightarrow\rho\gamma)$ & $3.0\pm1.0$ & $3.0\pm1.0$ \\
\hline
\end{tabular}
\end{center}
\caption[]{Fitted values of $\beta_A$, $\beta_B$ and $\beta_C$
together with the predicted values of 
the decay widths of 
$V \rightarrow S + \gamma$ and $S \rightarrow V + \gamma$
for $\theta_S \simeq-20^\circ$.
Only two out of four sets of $(\beta_A,\beta_B,\beta_C)$ are listed
here.
Units of $\beta_A$, $\beta_B$ and $\beta_C$ are $\mbox{GeV}^{-1}$
and those of the decay widths are keV.
}\label{tab1}
\end{table}

Let me make an analysis when the scalar mixing angle is 
taken as $\theta_S \simeq -90^{\circ}$.~\footnote{
 It should be noticed that 
 the predicted value of $f_0$-$\pi$-$\pi$ coupling
 for $\theta_S \simeq -90^\circ$
 is much larger than the value obtained in 
 Ref.~\cite{Harada-Sannino-Schechter:96}
 by fitting to the $\pi\pi$ scattering 
 amplitude~\cite{Black-Fariborz-Sannino-Schechter:99}
 as I discussed in section~\ref{sec:ELSM}.
}
As I stressed above, the values of $\beta_A$ and $\beta_C$ are
independent of the scalar mixing angle $\theta_S$.
The value of $\beta_B$ determined from 
$\Gamma(f_0\rightarrow\gamma\gamma)$ becomes
\begin{equation}
\beta_B = 
(1.1\pm0.1 \,,\, 0.12\pm0.13)\,\mbox{GeV}^{-1} \ .
\label{valB90}
\end{equation}
Then the typical predictions for $\Gamma(f_0\rightarrow\omega\gamma)$
and $\Gamma(f_0\rightarrow\rho\gamma)$ are given by
\begin{eqnarray}
&& \Gamma(f_0\rightarrow\omega\gamma) = ( 86\pm16)\,\mbox{keV}
\ ,
\nonumber\\
&& \Gamma(f_0\rightarrow\rho\gamma) = ( 3.4\pm3.2)\,\mbox{keV}
\ .
\end{eqnarray}
These predictions are very close to the ones in 
Eq.~(\ref{pred f0}).
This can be understood by the following consideration:
{}From the expression of $D_{f_0}^{\omega}$ in 
Eq.~(\ref{svgamps}),
one can see that it is 
dominated by the term including $\beta_C$
which is proportional to
$(\cos\theta_S - \sqrt{2} \sin\theta_S)$.
Then, the approximate relation
\begin{eqnarray}
&&
\cos( - 20^{\circ} ) - \sqrt{2} \, \sin( - 20^{\circ} )
\nonumber\\
&& \qquad
\simeq
\cos( - 90^{\circ} ) - \sqrt{2} \, \sin( - 90^{\circ} )
\simeq 1.4 
\label{eq20-90}
\end{eqnarray}
implies that
the value of $D_{f_0}^{\omega}$ for $\theta_S=-90^{\circ}$
is close to that for $\theta_S=-20^{\circ}$, and thus
$\Gamma(f_0\rightarrow\omega\gamma)$ for $\theta_S=-90^{\circ}$
to that for $\theta_S=-20^{\circ}$.
As for $\Gamma(f_0\rightarrow\rho\gamma)$ I should note that
the following relation is satisfied for
$X_{f_0}$ in Eq.~(\ref{samps}) 
and $D_{f_0}^{\rho^0}$ in Eq.~(\ref{svgamps}):
\begin{equation}
3 X_{f_0} - 2 \sqrt{2} D_{f_0}^{\rho^0}
= - \frac{4}{3} \sqrt{2} \beta_A ( c - \sqrt{2} s )
\ .
\end{equation}
Since the experimental value of 
$\Gamma(f_0\rightarrow\gamma\gamma)$, i.e., $X_{f_0}$ is used
as
an input, this relation implies that
the predicted value of $\Gamma(f_0\rightarrow\rho\gamma)$
for $\theta_S=-90^{\circ}$ is roughly equal
to that for $\theta_S=-20^{\circ}$.
Similarly, the predicted values of other radiative decay widths
for $\theta_S \simeq -90^\circ$
are also very close to those for $\theta_S \simeq -20^\circ$
as I list in Table~\ref{tab2}.
\begin{table}[htbp]
\begin{center}
\begin{tabular}{lrr}
\hline
$\beta_A$ & $0.72\pm0.12$ & $0.72\pm0.12$ \\
$\beta_B$ & $-0.12\pm0.13$ & $1.1\pm0.1$ \\
$\beta_C$ & $7.7\pm0.52$ & $7.7\pm0.52$ \\
\hline
$\Gamma(\sigma\rightarrow\gamma\gamma)$ 
   & $0.023\pm0.024$ & $0.37\pm0.10$ \\
$\Gamma(\phi\rightarrow\sigma\gamma)$ & $140\pm22$ & $35\pm11$ \\
$\Gamma(\omega\rightarrow\sigma\gamma)$ & $17\pm4$ & $33\pm5$ \\
$\Gamma(\rho\rightarrow\sigma\gamma)$ & $0.20\pm0.43$ & $17\pm4$ \\
$\Gamma(f_0\rightarrow\omega\gamma)$ 
  & $125\pm19$ & $86\pm16$ \\
$\Gamma(f_0\rightarrow\rho\gamma)$ & $18\pm8$ & $3.4\pm3.2$ \\
$\Gamma(a_0\rightarrow\omega\gamma)$ & $641\pm87$ & $641\pm87$ \\
$\Gamma(a_0\rightarrow\rho\gamma)$ & $3.0\pm1.0$ & $3.0\pm1.0$ \\
\hline
\end{tabular}
\end{center}
\caption[]{Fitted values of $\beta_A$, $\beta_B$ and $\beta_C$
together with the predicted values of 
the decay widths of 
$V \rightarrow S + \gamma$ and $S \rightarrow V + \gamma$
for $\theta_S \simeq-90^\circ$.
Only two out of four sets of $(\beta_A,\beta_B,\beta_C)$ are listed
here.
Units of $\beta_A$, $\beta_B$ and $\beta_C$ are $\mbox{GeV}^{-1}$
and those of the decay widths are keV.
}\label{tab2}
\end{table}

The result here indicates that it is difficult to
distinguish two pictures just from radiative decays.
Of course, other radiative decays 
should be studied to get more
informations on the structure of the scalar mesons.
Furthermore, inclusion of the loop corrections may 
be important~\cite{radiative}.
However, there are still large uncertainties in
the experimental data which make the analysis harder.
So instead of the analysis which can be compared with
experiment, some theoretical analyses give a clue to
get more informations on the structure of the scalar mesons.

\section{$\pi$-$\pi$ S\lowercase{cattering in} 
L\lowercase{arge} $N_c$ QCD}
\label{sec:pps}

In this section, I briefly review our recent 
analysis~\cite{Harada-Sannino-Schechter:04} on the 
$\pi$-$\pi$ scattering in QCD with large $N_c$, where
$N_c$ is the number of colors.

\begin{figure}[htbp]
\includegraphics [width = 6.5cm] {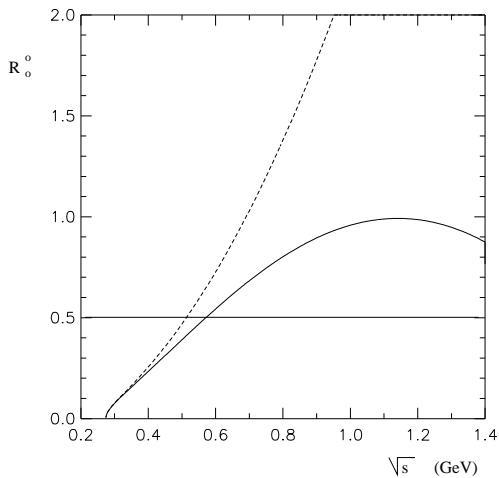}
\caption[]{%
Predicted curves for $R_0^0$~\cite{Harada-Sannino-Schechter:96}.
The horizontal axis shows the energy, and the vertical
axis shows the magnitude of the real part of the amplitude.
The dashed line shows the contribution from the
pion self interaction given by the current algebra.
The solid line shows the result with $\rho$-exchange
contribution included
in addition.
}\label{fig:pipi1}
\end{figure}

First, let me briefly review the analyses done in 
Refs.~\cite{Sannino-Schechter:95,Harada-Sannino-Schechter:96}
which stressed that the scalar meson $\sigma$ is needed
for satisfying the unitarity in the isospin $I=0$, $S$-wave
$\pi$-$\pi$ scattering amplitude
in real-life QCD with $N_c=3$.
First contribution included in the $\pi\pi$ scattering amplitude
is the one from the pion self interaction given by the current 
algebra, or equivalently, expressed by the leading order
chiral Lagrangian:
\begin{equation}
A_{\rm ca}(s,t,u) = \frac{ s - m_\pi^2 }{ F_\pi^2 }
\ ,
\label{ca amp}
\end{equation}
where $F_\pi = 92.42\,\mbox{MeV}$ is the pion decay constant.
The contribution from this to the real part of the $I=0$, $S$-wave
$\pi\pi$ scattering amplitude is shown by the dashed line
in Fig.~\ref{fig:pipi1}.
Since the amplitude greater than $0.5$ implies that the 
unitarity is violated, this amplitude breaks the
unitarity in the energy region around $500$\,MeV.
The solid line in Fig.~\ref{fig:pipi1} shows the curve
when the following $\rho$-exchange contribution is inclded
in addition:
\begin{eqnarray}
\lefteqn{
 A_{\rho}(s,t,u) =
 \frac{g_{\rho\pi\pi}^2}{2m^2_{\rho}}(4 m_{\pi}^2 - 3 s)
}
\nonumber \\
&&{}- \frac{g_{\rho\pi\pi}^2}{2}
\left[\frac{u-s}{(m^2_{\rho}-t)-im_{\rho}{\Gamma}_{\rho}\theta (t - 4
m_{\pi}^2)} \right.\nonumber \\
&&{}+ \left.
\frac{t-s}{(m^2_{\rho}-u)-im_{\rho}{\Gamma}_{\rho}\theta (u - 4
m_{\pi}^2)}\right]\ .
\label{rho contribution}
\end{eqnarray}
Note that the appearance of the first term is required by
the chiral symmetry.
{}From Fig.~\ref{fig:pipi1}, we can easily see that
a large cancellation occurs between the contribution
from the pion self-interaction 
and that from the $\rho$-meson exchange.
However, the unitarity is still violated around $550$\,MeV.

\begin{figure}[htbp]
\includegraphics [width = 6.5cm] {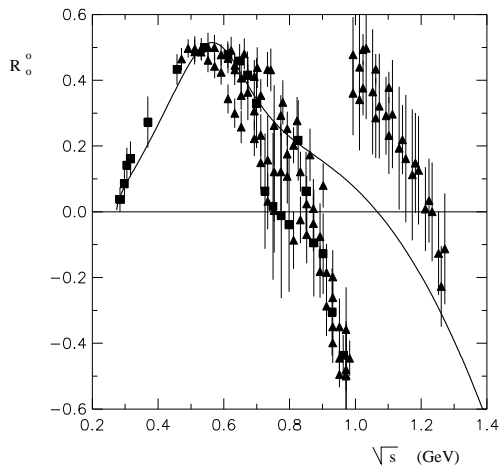}
\caption[]{%
Predicted curves for $R_0^0$
with the (current algebra) $+ \rho + \sigma$ contribution
included~\cite{Harada-Sannino-Schechter:96}.
}\label{fig:pipi2}
\end{figure}

To recover unitarity, we need negative contribution to the real part
above the point where the solid line in Fig.~\ref{fig:pipi1} violate
the unitarity.
While below the point a positive contribution is
preferred by the experiment.
Such property matches with the real part of a resonance contribution:
The resonance contribution is 
positive in the energy region below its
mass, while it is negative in the energy region above its mass.
In other words, the unitarity requires the existence of the
resonance in this energy region.
Then we have included a low mass broad scalar state, $\sigma$.
The contribution of the $\sigma$ to the real part of the amplitude 
is given by
\begin{equation}
\mbox{Re} A_{\sigma}(s,t,u)=
\frac{\gamma_{\sigma\pi\pi}^2}{2}
\frac{(s-2m_\pi^2)^2}{M^3_\sigma}
\frac{(M_\sigma^2-s)}{(s-M_\sigma^2)^2 +
M_\sigma^2{G^\prime}^2}\ ,
\label{eq:sigma}
\end{equation}
where $G^{\prime}$ is a parameter corresponding to the width
and 
$\gamma_{\sigma\pi\pi}$ is the $\sigma$-$\pi$-$\pi$ coupling 
constant.
This $\gamma_{\sigma\pi\pi}$ 
is related to the parameters $A$ and $B$ in 
the scalar-pseudoscalar-pseudoscalar interaction Lagrangian
given in Eq.~(\ref{NPP Lag}) 
as~\cite{Black-Fariborz-Sannino-Schechter:99}
\begin{equation}
\gamma_{\sigma\pi\pi} = 2 B \sin \theta_s 
  - 2 \sqrt{2} (B - A) \cos\theta_s
\ .
\end{equation}
In Ref.~\cite{Harada-Sannino-Schechter:96}, a best overall fit was
obtained with the parameter choices:
\begin{equation}
M_\sigma = 559\,\mbox{MeV}\ , \quad 
G^\prime = 370 \, \mbox{MeV} \, \quad
\gamma_{\sigma\pi\pi} = 7.8 \, \mbox{GeV}^{-1}\ .
\end{equation}
The result for the real part $R_0^0$ due to the inclusion of the
$\sigma$ contribution along with $\pi$ and $\rho$ contributions is
shown in Fig.~\ref{fig:pipi2}.
It is seen that the unitarity bound is satisfied and there is a
reasonable agreement with the experimental points up to about
$800$\,MeV.

The above analysis on the $\pi$-$\pi$ scattering in real-life QCD
tells an important lesson:
The mass of $\sigma$ meson is determined
by the point where the amplitude constructed from
$\pi + \rho$ contribution violates
the unitarity.

Now, let me show the results in
Ref.~\cite{Harada-Sannino-Schechter:04},
where the $\pi$-$\pi$ scattering in the large $N_c$ QCD was analyzed.

\begin{figure}[htbp]
\includegraphics [width = 6.5cm]{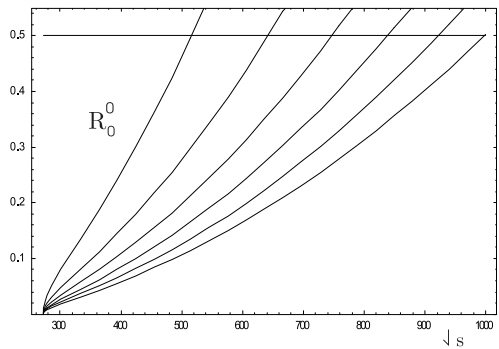}
\caption[]{%
Real part of the $I=0$ $S$-wave amplitude due to
the current algebra term plotted for the following increasing
values of $N_c$ (from left to right), 
$3$, $5$, $7$, $9$, $11$, $13$.~\cite{Harada-Sannino-Schechter:04}
}\label{fig:pipiNc1}
\end{figure}

First one to be included in the amplitude is 
the current algebra contribution given in Eq.~(\ref{ca amp}).
Note that the pion decay constant $F_{\pi}$ depends to leading order
on $N_c$ as $F_{\pi}(N_c)/F_\pi(N_c=3)=\sqrt{N_c/3}$,
while the pion mass
$m_{\pi}$ is independent of $N_c$ to leading order.
In Fig.~\ref{fig:pipiNc1},
I show the plot~\cite{Harada-Sannino-Schechter:04}
of the current algebra contribution to the real part of the $I=0$
$S$-wave amplitude, $R^0_0$
for increasing values of $N_c$. 
We observe that the unitarity is
violated at $s = s^{\ast}_{ca}$ which increases linearly
with $N_c$.

\begin{figure}[htbp]
\includegraphics [width = 6.5cm]{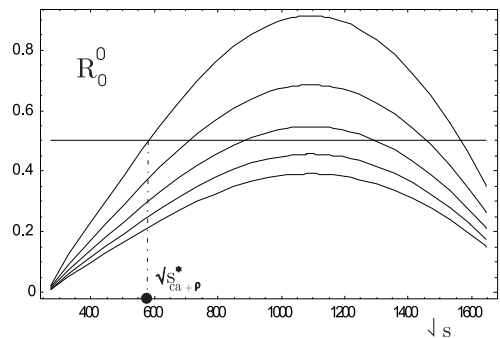}
\caption[]{%
Real part of the $I=0$ $S$-wave amplitude due
to the current algebra $+\rho$ terms, plotted for the following
increasing values of $N_c$ (from up to down),
$3$, $4$, $5$, $6$, $7$. The curve with largest
violation of the unitarity corresponds to $N_c=3$ while the
ones within the unitarity bound are for 
$N_c=6,7$.~\cite{Harada-Sannino-Schechter:04}
}\label{fig:pipiNc2}
\end{figure}

Next, I show that this result is strongly
modified by the presence of the well established $q\bar{q}$
companion of the pion -- the $\rho$ meson.
The amplitude is obtained by adding to the current algebra
contribution the $\rho$ meson contribution given in 
Eq.~(\ref{rho contribution}).
In Fig.~\ref{fig:pipiNc2}, I show the plot of
$R^0_0$ due to current algebra plus the $\rho$
contribution for increasing values of $N_c$. 
Here the scaling property of the $\rho$-$\pi$-$\pi$ coupling
is taken as
$g_{\rho\pi\pi}(N_c)/g_{\rho\pi\pi}(N_c=3) = \sqrt{3/N_c}$
with $m_\rho$ kept fixed.
This figure shows 
that the unitarity (i.e., $\vert R_0^0 \vert \le 1/2$)
is satisfied for $N_c\geq 6$ till well beyond
the 1~GeV region. However unitarity is still a problem for $3$, $4$
and $5$ colors.

\begin{figure}[htbp]
\includegraphics [width = 6.5cm]{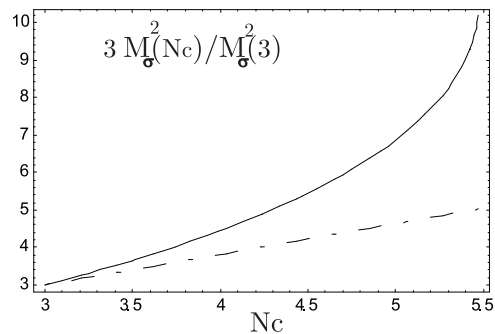}
\caption[]{%
The value $M^2(N_c)$ as a function of $N_c$ and normalized to 
$M^2(3)/3$.~\cite{Harada-Sannino-Schechter:04}
The
dashed line corresponds to the pure current algebra contribution
while the solid line to current algebra $+\rho$ contribution.
}\label{Massemp}
\end{figure}

As I showed in Fig.~\ref{fig:pipi2},
the violation of the unitarity in the real-life QCD is recovered 
by the existence of the $\sigma$ pole.
The $\sigma$ pole structure is such that the
real part of its amplitude is positive for $s<M^2_{\sigma}$ and
negative for $s>M^2_{\sigma}$. 
Identifying the squared sigma mass roughly
with $s^{\ast}$, at which $R_0^0$ without $\sigma$ contribution
violates unitarity,
will then give a negative contribution where
the real part of the amplitude exceeds $+0.5$. In the
case when only the current algebra term is included we get
\begin{eqnarray}
M^2_{\sigma} \approx s^{\ast}_{ca}=4\pi\,F^2_{\pi} \ .
\end{eqnarray}
This shows that the squared mass of the $\sigma$ meson
needed to restore unitarity for $N_c=3,4,5$ increases
 roughly linearly with $N_c$. 
This estimate gets modified a bit
when we include the vector meson (see Fig.~{\ref{Massemp}}),
yielding
$M^2_{\sigma} \approx
s^{\ast}_{ca+\rho}$, where $s^{\ast}_{ca+\rho}$ is to be obtained
from Fig.~\ref{fig:pipiNc2}.
This clearly shows that the mass of $\sigma$ becomes larger for
larger $N_c$,
and when $N_c \ge 6$,
the $\sigma$ is not needed in the energy region below 2 GeV.
{}From this we concluded~\cite{Harada-Sannino-Schechter:04}
that the $\sigma$ meson is unlikely the 2-quark state and
likely the 4-quark state.
This is similar to the conclusion obtained in 
Ref.~\cite{largeNc:pipi}.

\section{S\lowercase{ummary}}
\label{sec:sum}

In this write-up, focusing on the structure of the
low-lying scalar nonet,
I summarized the analyses in 
two works~\cite{BHS,Harada-Sannino-Schechter:04}
which I have done recently.

In section~\ref{sec:ELSM}, 
following 
Refs.~\cite{Black-Fariborz-Sannino-Schechter:99,Fariborz-Schechter},
I first briefly reviewed
what the hadronic processes involving the scalar nonet tell about
the quark structure of the low-lying scalar mesons.
The analysis on the pattern of the hadronic processes
implies that the scalar nonet takes dominantly
the $qq\bar{q}\bar{q}$ structure 
(or the diquark--anti-diquark structure).
Next, in section~\ref{sec:rad},
I summarized the work in Ref.~\cite{BHS}, in which
we analyzed the radiative decays involving the 
scalar nonet based on the $qq\bar{q}\bar{q}$ picture.
I also presented a new result~\cite{BHS:prep} on the
radiative decays based on the $q\bar{q}$ picture.
Our result indicates that it is difficult to
distinguish two pictures just from radiative decays.
Finally, in section~\ref{sec:pps}, I summarized the work in
Ref.~\cite{Harada-Sannino-Schechter:04}, in which we 
studied the $\pi$-$\pi$ scattering in large $N_c$ QCD.
Our analysis shows that the mass of the $\sigma$ meson
becomes larger for larger $N_c$, and when $N_c\ge6$,
the $\pi$-$\pi$ scattering amplitude satisfies the unitarity
without the $\sigma$ meson.
{}From this we concluded that the $\sigma$ meson is unlikely the 
$q\bar{q}$ state and likely the $qq\bar{q}\bar{q}$ state.

\section*{Acknowledgments}
I would like to thank the organizers to give an opportunity
to present my talk.
I am grateful to Deirdre Black, Francesco Sannino and
Joe Schechter for collaboration in 
Refs.~\cite{BHS,Harada-Sannino-Schechter:04} on which this 
talk is based.
My work is supported in part by
the JSPS Grant-in-Aid for Scientific Research (c) (2) 16540241,
and by 
the 21st Century COE
Program of Nagoya University provided by Japan Society for the
Promotion of Science (15COEG01).

\end{document}